\documentstyle[12pt]{article}
  
\def\bo{{\raise.15ex\hbox{\large$\Box$}}}

\def\dag{^{\dagger}{}}

\def\ordless{{\lower2mm\hbox{$\,\stackrel{\textstyle <}{\sim}\, $}}}
\def\ordmore{{\lower2mm\hbox{$\,\stackrel{\textstyle >}{\sim}\, $}}}

\newtoks\slashfraction
\slashfraction={.13}
\def\slash#1{\setbox0\hbox{$\, #1$}
\setbox0\hbox to \the\slashfraction\wd0{\hss \box0}/\box0}

\def\leftrightarrowfill{$\mathsurround=0pt \mathord\leftarrow \mkern-6mu
        \cleaders\hbox{$\mkern-2mu \mathord- \mkern-2mu$}\hfill
        \mkern-6mu \mathord\rightarrow$}
\def\overleftrightarrow#1{\vbox{\ialign{##\crcr
        \leftrightarrowfill\crcr\noalign{\kern-1pt\nointerlineskip}
        $\hfil\displaystyle{#1}\hfil$\crcr}}}
\def\startarray{\left( \begin{array}}
\def\finarray{\end{array} \right)}
\def\starteq{
\begin{eqnarray}}
\def\fineq{\end{eqnarray}
}

\catcode`@=11
\def\underline#1{\relax\ifmmode\@@underline#1\else
$\@@underline{\hbox{#1}}$\relax\fi}
\catcode`@=12
 
\newskip\humongous \humongous=0pt plus 1000pt minus 1000pt

\newif\ifdtup
 
\def\textcite#1{Ref.~{\cite{#1}}}

\def\thefootnote{\fnsymbol{footnote}}
 
\def\author#1#2{{\bf #1} \\ {\em #2}\vspace{5mm}}

\def\bold#1{\setbox0=\hbox{$#1$}%
     \kern-.025em\copy0\kern-\wd0
     \kern.05em\copy0\kern-\wd0
     \kern-.025em\raise.0433em\box0 }

\def\title#1#2#3#4#5{\thispagestyle{empty}
        \begin{center} \vspace*{1cm} { \bf #3} \\[.5in] {#4{}}
        \end{center} \vfill \centerline{ ABSTRACT}
   {\nopagebreak \noindent\begin{quotation}\noindent {\small #5}
   \end{quotation}} \vfill {#2} \hfill\begin{tabular}{r} {#1} 
        \end{tabular}  \newpage
        \def\thefootnote{\arabic{footnote}}}
\def\prefer{\section*{}
    \list{[\arabic{enumi}]}{\usecounter{enumi}\settowidth\labelwidth{[000]}
      \leftmargin\labelwidth\advance\leftmargin\labelsep \rightmargin=0pt}
        \small \sfcode`\.=1000\relax}

\def\refer#1{\section*{\large \sc {#1}}
    \list{\arabic{enumi}.}{\usecounter{enumi}\settowidth\labelwidth{[000]}
      \leftmargin\labelwidth\advance\leftmargin\labelsep \rightmargin=0pt}
        \raggedright \small \sfcode`\.=1000\relax}

\def\ReFer#1#2{\section*{\large\sc#1}
    \list{[\arabic{enumi}]}{\usecounter{enumi}\settowidth\labelwidth{#2}
      \leftmargin\labelwidth\advance\leftmargin\labelsep \rightmargin=0pt}
        \raggedright \small \sfcode`\.=1000\relax}

\def\REFER#1#2{\section*{\large\sc#1}
    \list{#2 {enumi}.}{\usecounter{enumi}\settowidth\labelwidth{[000]}
      \leftmargin\labelwidth\advance\leftmargin\labelsep \rightmargin=0pt}
        \raggedright \small \sfcode`\.=1000\relax}

\def\startbib{\vspace{1in}\begin{refer}{References}
\small\frenchspacing\nopagebreak}
\def\endbib{\end{refer} \normalsize \nonfrenchspacing}
\def\startfig{\newpage \centerline{{\sl Figure captions}} \begin{itemize}}

\def\endfig{\end{itemize}}
 
\newcommand{\be}{\begin{equation}}
\newcommand{\ee}{\end{equation}}
\newcommand{\bea}{\begin{eqnarray}}
\newcommand{\eea}{\end{eqnarray}}

\newcommand{\AmS}{{\protect\the\textfont2
  A\kern-.1667em\lower.5ex\hbox{M}\kern-.125emS}}

\begin{document}
\title{September 1999} {PC738.0999}
{ CP-VIOLATING ASYMMETRY IN $B$ DECAYS TO \\ 3 PSEUDOSCALAR MESONS 
\footnotetext{${\dag}$  
Talk given at the QCD Euroconference 99, Montpellier 7-13 July 1999}}
{\author{T. N. Pham} {Centre de Physique Th\'eorique, \\
Centre National de la Recherche Scientifique, UMR 7644, \\  
Ecole Polytechnique, 91128 Palaiseau Cedex, France}}        
{The measurement  of CP asymmetries in charged B  meson decays  might provide 
the first demonstration 
of CP violation outside the K system.
Among the usual three CP-odd phases $\alpha$, $\beta$ and $\gamma$, 
the  phase $\gamma$ 
seems to be the most difficult to explore experimentally.
In this talk I would like to report on a recent analysis  of 
  the CP asymmetry  in the partial widths for the non-leptonic decays 
$B^{\pm} \rightarrow M {\bar M} \pi^{\pm}$ 
($ M = \pi^+, K ^+ , \pi^0, \eta$), which results from the
interference of
the non-resonant decay amplitude with the resonant amplitude for
$B^{\pm} \rightarrow \chi_{c0} \pi^{\pm} $
followed by the decay $\chi_{c0} \rightarrow M {\bar M} $.  The CP violating phase
$\gamma$ can be extracted from the measured asymmetry.
We find that the partial width asymmetry for $B^\pm \rightarrow \pi^+ \pi^- \pi^\pm$
is about $0.33~\sin \gamma$, and about $0.45~ \sin \gamma$ for 
$B^\pm \rightarrow K^+ K^-\pi^\pm$, while it is somewhat smaller for 
$B^\pm \rightarrow \pi^0 \pi^0 \pi^\pm$ and $B^\pm \rightarrow \eta \eta \pi^\pm$.}

The latest values for  $\epsilon'/\epsilon$ measured recently by the 
NA31 \cite{Peyaud} and
the KTeV experiments \cite{Hsiung} show clearly that 
there is a $\Delta S=1$ direct
CP-violating non-leptonic weak interaction in $K \rightarrow \pi\pi$
decays. This brings us closer than ever to the Cabibbo-Kobayashi-Maskawa
(CKM) model as a 
description of CP violation in the standard model. 

In the standard model,each generation of quark or 
lepton couples to the $W^{\pm}$ gauge boson
through a left-handed doublet. If there is no mixing between fermions of
different generation,there would be no $\Delta S=1$ charged 
current. Hyperons and kaons would not decay into lighter hadrons by weak
interactions. Thus mixings between quarks of different generation must 
occur to account for the  $\Delta S =1$ 
and $\Delta B =1$ charged currents.The mixing would probably come from
the flavor-changing mixing in the quark mass matrix. After the rotation
of the quark fields by the unitary CKM quark mixing matrix to eliminate the
non-diagonal flavor-changing quark mass terms, the weak interaction
eigenstates differ from the quark mass eigenstates by this unitary
transformation so that, in terms of the mass eigenstates,the weak 
charged currents will now contain  flavor-changing terms.
Since the neutral current is not affected by the unitary transformation
on the quark fields,
flavor-changing neutral current is absent at the tree level as implied
by the GIM mechanism \cite{Glashow}. This GIM mechanism already 
tells us that the quark mixing matrix must be unitary. Furthermore,
with three generations, CP violation  can  be
generated by the phase of the elements of $V$ \cite{KM}. Thus the
weak interaction charged
currents are generated with a minimum number of parameters. 
The CKM quark mixing matrix can thus be considered
as a generalised universality for the weak interactions. V is usually 
defined as 

\be
\pmatrix{d' \cr s' \cr b'}= \pmatrix{V_{ud}& V_{us} & V_{ub} \cr
V_{cd} & V_{cs} & V_{cb} \cr 
V_{td} & V_{ts} & V_{tb}} \pmatrix{d \cr s \cr
b}
\label{Vckm} 
\ee
where $d,s,b$ and $d',s',b'$ are respectively the  mass eigenstates
and weak interaction eigenstates for the charge $Q=-1/3$ quarks.
Unitarity of $V$ implies that any non-diagonal element of $VV^{\dag}$
is zero. For example, for the  $(db)$ elements relevant to $B$ decays,
we have: 
\be
V_{ud}V_{ub}^{*} + V_{cd}V_{cb}^{*} + V_{td}V_{tb}^{*} =0 
\label{db} 
\ee
This can be represented by a triangle \cite{PDG} with the three 
angles $\alpha$, $\beta$
and $\gamma$ expressed in terms of the CKM matrix elements as:
\bea
&& \kern -0.5cm \alpha = arg(-V_{td}V_{tb}^{*}/V_{ud}V_{ub}^{*}) \\ \nonumber
&& \kern -0.5cm \beta = arg(-V_{cd}V_{cb}^{*}/V_{td}V_{tb}^{*}) \\ \nonumber
&& \kern -0.5cm \gamma = arg(-V_{cd}V_{cb}^{*}/V_{ud}V_{ub}^{*})
\label{arg}
\eea
The area of the triangle is given by:
\be
A({\rm db}) = (1/2){\rm Im}\,\left(V_{cd}V_{cb}^{*}V_{ud}V_{ub}^{*} \right)
\label{area} 
\ee
which appears 
in the interference term between the CP-conserving and 
CP-violating physical amplitude and is a measure of CP
violation. If one of the terms in (\ref{db}) is zero or real,
the interference term
vanishes since the other two terms are relatively real \cite{Bigi} . 
This shows that it is  not be possible to generate CP violation 
effect by the quark mixing matrix if there are only 2 generations
of quark. Since  the 3 sides of the (db)-unitarity triangle
are of comparable length, the 3 angles $\alpha$, $\beta$ and $\gamma$
are also comparable and large, CP violations in $B$ decays and 
$B^{0}-\bar{B}^{0}$ mixing would be appreciable and a great deal of
efforts are being devoted to the measurements of the CP violations in
$B$ decays which will allow us to test the validity of the CKM model.
By measuring the three angles, we can see if they all add up to $\pi$ 
as predicted by the CKM model. The angle $\alpha$ and $\beta$ can be
determined with good accuracy from the $B \rightarrow \pi\pi$ and 
$B \rightarrow \Psi K_{S}$. The measurement of $\gamma$
 in $B_{s}\rightarrow \rho^{0}K_{S}$ 
is more challenging because of the rapid $B_{s}-\bar{B}_{s}$
oscilations \cite{Bigi}. Other direct CP violation measurements of
$\gamma$ may be possible, as first pointed out in
\cite{Deshpande,Eilam}. For example,
in $B \rightarrow 3\pi$ decays,the two pions in the final state
can have a large invariant mass around 
the charmonium  $\chi_{c0}$ state at $34 17\,\rm MeV$
which  decays into an $S$-wave two-pion state and hence an interference
occurs between the non-resonant and the resonant amplitude coming from
the decay of the charmonium state. If the two amplitudes are comparable, 
CP asymmtry could be large and a direct CP violation in $B \rightarrow 3\pi$
and  $B \rightarrow K \bar{K}\pi$ decays could be measured. This looks 
quite feasible as the theoretical branching ratio which is  
$1.5\times 10^{-5}$ to $8.4\times 10^{-5}$ as given in \cite{Deshpande} 
and the recent CLEO upper limits \cite{CLEO}  
$\rm BR(B^+ \rightarrow \pi^+ \pi^- \pi^+) \le 4.1 \times 10^{-5}$ and
$\rm BR(B^+ \rightarrow K^+ K^-\pi^+) = 7.5 \times 10^{-5}$ indicate that these 
3-body $B$ decays  could be measured soon. In this talk, I
would like to report on a recent work \cite{Fajfer} on the $B$ decays 
into 3 pseudoscalar mesons as a possible way to see direct CP violation
in B decays and to measure the angle $\gamma$. In the following, I
present only the main results of our work, as the details can be found
in this reference.

Consider now the non-resonant $B^{\pm} \rightarrow M\bar{M}\pi^{\pm}$, 
$M = \pi^{+},K^{+},\pi^{0},\eta$ decays.
The weak effective Lagrangian for the  Cabibbo-
suppressed non-leptonic $B$  decays is given by
\begin{equation}
{\cal L}_{w} = - \frac{G_F}{{\sqrt 2}} V_{ud}^* V_{ub} (a_1^{eff} O_1 +
a_2^{eff} O_2)
\label{nonleptonic}
\end{equation}
where  $O_1 = ({\bar u} b)_{V-A}\, ({\bar d} u)_{V-A} $  and
$O_2 = ({\bar u} u)_{V-A}\, ({\bar d} b)_{V-A} $ with
$a_1^{\rm eff}  \simeq 1.08$, $a_2^{eff} \simeq 0.21$ their effective
short-distance coefficients taken from fits to 
two-body $B$ decays \cite{Browder}. Contributions from penguin operators
are expected to be small and are neglected here. 
To obtain the decay amplitude,we use
the factorisation approach as in \cite{Deshpande}. The matrix
elements $<\kern -0.1667cm M\bar{M}\pi|{\cal L}_{w}|B \kern -0.1667cm>$ 
can thus be written as a
product of two $(V-A)$ current matrix elements taken in all possible ways
between the initial and final state. The annihilation term 
$<\kern -0.1667cm M\bar{M}\pi|({\bar d} u)_{A}|0 \kern -0.1667cm>\break
<\kern -0.1667cm 0|({\bar u} b)_{A}|B \kern -0.1667cm>$ is 
$O(m_{\pi}^{2})$ and is negligible. The $V\times V$ term 
$<\kern -0.1667cm M\bar{M}|({\bar u} u)_{V}
|0\kern -0.1667cm><\kern -0.2cm\pi|({\bar d} b)_{V}|B\kern -0.1667cm>$ 
from $O_{2}$, is dominated
by the $\rho$ resonance, but is relatively small compared to the $O_{1}$  
contribution and is also suppressed in the region of large pion momentum
and is therefore not relevant to our analysis of the CP asymmetry. So
the most important term in the decay amplitude comes from 
terms of the form\break
$<\kern -0.1667cm M\bar{M}|({\bar u} b)_{V-A}|B^{-}\kern -0.1667cm>
<\kern -0.1667cm\pi^{-}|({\bar d} u)_{A}|0\kern -0.1667cm>$ which 
contains the $B^{-}\rightarrow M\bar{M}l\bar{\nu}$ semi-leptonic decay
form factors. We obtain these $B_{l4}$ form factors 
by extrapolating the results 
for $D^{+} \rightarrow K^{-}\pi^{+}l\nu$ \cite{Bajc}
to the $B$ meson using scaling law in the heavy quark 
limit \cite{Casalbuoni}. These form factors are given as \cite{Lee}
\starteq
 <\kern -0.1667cm \pi^- (p_1) \pi^+ (p_2) | {\bar u} 
\gamma_{\mu} (1 - \gamma_5) b | B^- (p_B) \kern -0.1667cm>
\kern -0.2cm  &=&\kern -0.2cm
 i\,r(p_B-p_2-p_1)_\mu + i\,w_+(p_2+p_1)_\mu\nonumber\\
&&\kern -0.4cm {\hfill} + i\,w_-(p_2-p_1)_\mu -  
2\,h\,\epsilon_{\mu\alpha\beta\gamma}p_B^\alpha p_2^\beta p_1^\gamma\;.
\label{wwh}
\fineq
The non-resonant $B \rightarrow M\bar{M}\pi$ decay amplitudes
can then be obtained in terms of these 
form factors in a straightforward manner.
I give here the expressions for the $B^- \rightarrow \pi^+ \pi^- \pi^-$ 
amplitude as the other amplitudes can be found in \cite{Fajfer}. We
have then
\starteq
 {\cal M}_{nr} (B^-  \rightarrow \pi^-  \pi^- \pi^+ ) 
\kern -0.2cm  &=&\kern -0.2cm
\frac{G_F}{{\sqrt 2}} V_{ud}^* V_{ub}\times 
  \{a_1^{eff} [ {f_{\pi} \over 2} 
(m_B^2 -s -m_{\pi}^2) w_+^{nr} (s,t) +\nonumber \\ 
&& \kern -0.4cm {\hfill} {f_{\pi} \over 2} (2 t + s - m_B^2 - 3 \,m_{\pi}^2) w_-^{nr}(t) ]
 +   (s \leftrightarrow t)\}, 
\label{amplitude}
\fineq
\ where  
\starteq
\kern -0.5cm w_+^{nr}(s,t) \kern -0.2cm  &=&\kern -0.2cm - \frac{g}{f_{\pi}^2} 
\frac{f_{B*} m_{B*}^{3/2} m_B^{1/2}}{t - m_{B*}^2} \times 
 [ 1 - \frac{1}{2 m_{B*}^2} 
(m_B^2 -m_{\pi}^2 - t) ] + \frac{f_B}{ 2 f_{\pi}^2} \nonumber\\
&&  \kern -0.4cm {\hfill} - \frac{{\sqrt m_B} 
\alpha_2}{ 2 f_{\pi}^2} 
\frac{1}{m_B^2}(2 t + s - m_B^2 - 3 \,m_{\pi}^2 ), 
\label{w+1}
\fineq
and
\starteq
 w_-^{nr}(t)  =   \frac{g}{f_{\pi}^2} 
\frac{f_{B*} m_{B*}^{3/2} m_B^{1/2}}{t - m_{B*}^2} \times 
[ 1 + \frac{1}{2 m_{B*}^2}(m_B^2 -m_{\pi}^2 - t)] 
 +\frac{{\sqrt m_B} \alpha_1}{  f_{\pi}^2}.    
\label{w-1}
\end{eqnarray}
where $g$ is the $B^{*}B\pi$ coupling constant. $\alpha_{1,2}$ are
two parameters for the direct terms obtained with chiral perturbation
theory \cite{Bajc}. The amplitudes given in \cite{Deshpande} are obtained
with $B^{*}$ pole dominance and no direct terms. However, we found that,
although the $B^{*}$ pole terms are important in the region
with large $t$, for small $t$, the direct terms become appreciable and
comparable to the $B^{*}$ pole terms.

With  $f_B \simeq 128\,\rm MeV$ obtained from $f_D \simeq 200\,\rm MeV$ 
using the relation \cite{Casalbuoni},
\be
f_{B} = \sqrt{m_{D}/m_{B}}
\ee
The coupling constant $g$, which is independent of the heavy quark mass, 
can also be obtained from the $D^{*}\rightarrow D\pi$ which gives 
$g = 0.3 \pm 0.1$ \cite{PC}. This is more or less consistent with the
values $g = 0.15 \pm 0.08$ \cite{Bajc1} obtained from $D^0 \rightarrow K^- l^+ \nu$
decays. However, we find that  for $g$ in the range
$0.2 \leq g \leq 0.23$ and  $\alpha_1^{B\rho} = -0.13$ $\, \rm GeV^{1/2}$, 
$\alpha_2^{B\rho} = -0.40$ $\, \rm GeV^{1/2}$ obtained by extrapolating
from the corresponding $D \rightarrow K^{*}$ values, a branching ratio
consistent with the experimental upper limits \cite{CLEO}
can be obtained. \kern -0.2cm We find: 
$3.4 \times 10^{-5}\leq$ $BR(B^- \rightarrow \pi^- \pi^+ \pi^+) \leq$
 $ 3.8 \times 10^{-5}$, 
$1.4 \times 10^{-5}\leq$$BR(B^-\rightarrow K^- K^+ \pi^-) \leq$ $1.5 \times 10^{-5}$,
We also obtained using the same set of parameters~:
$1.5 \times 10^{-5}\leq$
$BR(B^- \rightarrow \pi^- \pi^0 \pi^0)\leq$ $ 1.7 \times 10^{-5}$ and  
$1.0 \times 10^{-5}\leq$
$BR(B^- \rightarrow \pi^- \eta \eta)\leq$ $ 1.1\times 10^{-5}$ . We note that 
the contributions to the branching ratios from $\alpha_{1,2}$  
are very important and that the upper and
lower limit we give above correspond to the two values $g=0.23$ and
$g=0.20$ respectively. The
difference is insignificant and one can say that to produce a branching
ratio consistent with the experimantal limit, $g$ should be around 0.20.

The resonant decay amplitude, in the narrow width approximation, is
given by \cite{Deshpande,Eilam}~:
\starteq
 {\cal M}_{r}(B^- \rightarrow  \chi_{0c} \pi^- \rightarrow M \bar M
 \pi^-)
\kern -0.2cm  &=&\kern -0.2cm  M(B^{-} \rightarrow \chi_{0c} \pi^- ) 
\frac{1}{ s - m_{\chi_{0c}}^2 + 
i \Gamma_{\chi_{0c}} m_{\chi_{0c}}} \times \nonumber\\
 &&\kern -0.4cm {\hfill} M( \chi_{0c} \rightarrow M \bar M ) 
+ (s \leftrightarrow t).
\label{resonant}
\fineq
To have an estimate for the resonant amplitude, we use the estimate
$BR(B^{\pm} \rightarrow $ $\chi_{c0}$ $ \pi^{\pm} )\times$ $ BR(\chi_{c0} $ 
$\rightarrow \pi^+ \pi^-)  =$ $ 5 \times 10^{-7}$ derived in
\cite{Deshpande,Eilam} and the $\chi_{c0}$ decay data \cite{PDG1}.
The interference term which produces CP asymmetry occurs 
in the kinematical region where the $M\bar M$ invariant mass 
is close to the mass of the $\chi_{c0}$ meson. The
partial width in this region where s is between 
$ s_{min} = (m_{\chi_{0c}} - 2\Gamma_{\chi_{0c} })^2$ and
$ s_{max} = (m_{\chi_{0c}} + 2\Gamma_{\chi_{0c} })^2$, is given by:
\starteq
\Gamma_p\kern -0.2cm &=& \kern -0.2cm
\frac{1}{(2 \pi)^3} \frac{1}{32 m_B^3} \times 
\int_{s_{min}}^{s_{max}} ds \int_{t_{min}(s)}^{t_{max}(s)} 
dt~|{\cal M}_{nr } + {\cal M}_{r} |^2 . 
\label{partial}
\fineq
The CP-violating asymmetry is defined by 
\begin{equation}
 A = \frac{ \Gamma_p - \bar{\Gamma}_{p}}
{\Gamma_p + \bar{\Gamma}_{p}}.
\label{asym}
\end{equation} 
 
With  the same values for $g$ and $\alpha_{1,2}$ used in the total
decay rates calculation, we find, for the absolute of the asymmetry $|A|$;
\newpage
\bea
&& \kern -0.57cm 0.33~\sin\gamma \leq |A(B^\pm \rightarrow \pi^+ 
\pi^- \pi^\pm)| \leq  0.34~\sin\gamma \nonumber\\
&& \kern -0.57cm 0.44~\sin\gamma \leq |A(B^\pm \rightarrow K^+ K^- 
\pi^\pm)|  \leq  0.45~\sin \gamma,     \nonumber\\
&& \kern -0.57cm 0.23~\sin\gamma \leq |A(B^\pm \rightarrow \pi^0 
\pi^0 \pi^\pm)|  \leq  0.24~\sin\gamma \nonumber\\
&& \kern -0.57cm 0.17~\sin\gamma \leq |A(B^\pm \rightarrow \eta\, 
\eta \,\pi^\pm)|  \leq  0.20~\sin\gamma. 
\label{absolute}
\eea
As mentioned above, the upper and lower limit for the absolute
CP asymmetry $|A|$ in (\ref{absolute}) corresponds to $g=0.23$ and $g=0.20$
respectively. The difference in the two values is insignificant.

In conclusion, we have analyzed the partial width CP asymmetry
in $B^{\pm}\rightarrow M\bar M\pi^{\pm}$ decays ($M = \pi^+$, 
$K^+$, $\pi^0$,  $\eta$) . Depending on
the decay modes, the CP asymmetry is estimated to vary between
$0.2\,\sin\gamma$ and $0.45\,\sin\gamma$. These values could suffer from
various uncertainties coming from the approximation for the 3-body
amplitude and experimental errors in the CKM matrix elements and 
form factors involved, but would remain to be large so that direct
CP violation could provide us with a mean to measure
the angle $\gamma$.

\bigskip

I would like to thank S. Narison and the organisers of QCD99 for the
warm hospitality extended to me at Montpellier.

\end{document}